\documentclass[preprint,12pt]{elsarticle}

\usepackage{amssymb}
\usepackage{amsmath}
\usepackage{enumitem}
\usepackage{tabularx}
\usepackage{todonotes}

\usepackage{makecell}
\usepackage{hyperref}

\journal{Computer Speech and Language}

\begin{document}

\begin{frontmatter}



\title{Summary of the NOTSOFAR-1 Challenge: Highlights and Learnings}


\author[microsoft]{Igor Abramovski\fnref{1}} 
\author[microsoft]{Alon Vinnikov\fnref{1}} 
\author[microsoft]{Shalev Shaer} 
\author[microsoft]{Naoyuki Kanda}
\author[microsoft]{Xiaofei Wang}
\author[microsoft]{Amir Ivry} 
\author[microsoft]{Eyal Krupka} 

\fntext[1]{Equal contribution}

\affiliation[microsoft]{Microsoft}

\begin{abstract}
The first Natural Office Talkers in Settings of Far-field
Audio Recordings (NOTSOFAR-1) Challenge is a pivotal initiative that sets new benchmarks by offering datasets more representative of the needs of real-world business applications than those previously available. The challenge provides a unique combination of 315 recorded meetings across 30 diverse environments, capturing real-world acoustic conditions and conversational dynamics, and a 1000-hour simulated training dataset, synthesized with enhanced authenticity for real-world generalization, incorporating 15,000 real acoustic transfer functions. In this paper, we provide an overview of the systems submitted to the challenge and analyze the top-performing approaches, hypothesizing the factors behind their success. Additionally, we highlight promising directions left unexplored by participants. By presenting key findings and actionable insights, this work aims to drive further innovation and progress in DASR research and applications.
\end{abstract}




\end{frontmatter}



\section{Introduction}
\label{sec_intro}

With the rapid advancements in Artificial Intelligence (AI) and the increasing reliance on virtual communication tools, the field of distant speaker diarization and automatic speech recognition (DASR) has become more crucial than ever. The emergence of Large Language Models (LLMs) has further amplified the importance of DASR, enabling a wide range of applications such as meeting summaries, note-taking, sentiment analysis, and personalized, context-aware responses to user queries. These capabilities are particularly valuable in meeting scenarios, where accurate distant speech transcription and speaker diarization are essential for effective communication and collaboration.

Advancements in the DASR field have been long supported by comprehensive and high-quality datasets. For example, the AMI dataset \cite{carletta2005ami} provides 100 hours of real English meeting recordings with close-microphone annotations, which has been widely used as a benchmark for DASR. More recently, the CHiME-5/6 challenge \cite{barker2018chime5,watanabe2020chime6} provided about 50 hours of daily conversation recordings along with baseline DASR systems, significantly advancing DASR research. The LibriCSS dataset \cite{chen2020css} provides evaluation data on more controlled environments by playing LibriSpeech \cite{panayotov2015librispeech} utterances via loudspeakers. Nevertheless, there are still significant gaps between real-world complexity, such as limited recording environments (AMI), fixed numbers of speakers (AMI, CHiME, LibriCSS), etc. These gaps pose challenges to developing a DASR system that works under real-world uncertainty.

To address this, the Natural Office Talkers in Settings of Far-field Audio Recordings (NOTSOFAR-1) challenge \cite{vinnikov2024notsofar}, an international competition in DASR, was introduced at Interspeech 2024 as part of CHiME's annual challenge (CHiME-8 \cite{chime8}) \footnote{A more detailed comparison between NOTSOFAR-1 and other benchmarks can be found in \cite[Section 1]{vinnikov2024notsofar}}. It provided comprehensive datasets as well as a baseline system to facilitate research and development in the field of DASR.
The primary goals of the NOTSOFAR-1 challenge were to advance the state-of-the-art in DASR by offering new datasets that capture the complexities of distant meeting transcription and to provide a platform for researchers to evaluate their systems in realistic scenarios. The challenge aimed to bridge the gap between existing datasets and real-world applications, focusing on single-channel and known-geometry multi-channel tracks, which are commonly used in conference room environments. It introduced two new datasets: a benchmark dataset of 315 English meetings recorded across 30 different rooms, and a 1000-hour simulated training dataset carefully designed to generalize well to real-world scenarios. The simulated dataset incorporates 15,000 \emph{real} acoustic transfer functions (ATFs) \cite{habets2006room} and features high-quality clean speech, providing a rich resource for training speech separation and enhancement methods. 

In this paper, we present a comprehensive review of the results from the NOTSOFAR-1 challenge. We summarize and analyze the submitted works, highlighting the key findings and contributions of the participating teams.
We compare the structure of submitted pipelines and divide them into two classes: Dia-Sep-ASR and CSS-ASR-Dia (see \ref{sec_systems}). We compare the performance of each class of systems in both single-channel and multi-channel tracks and offer conclusions. We delve into the ablation studies of top-performing systems and hypothesize which components contributed most significantly to the success of the winning team. We compare the performance of single-channel and multi-channel systems and confirm a significant gap that single-channel systems are currently unable to bridge. We survey the neural net models participants experimented with, aiming to compete with or enhance the dominant guided source separation (GSS) \cite{boeddeker2018gss}. We find that many teams invested in fine-tuning their automatic speech recognition (ASR) components and survey their choice of models, fine-tuning approaches, and inference techniques. We identify and discuss unexplored areas we believe hold significant promise.

In what follows, we describe the NOTSOFAR-1 Challenge in more details (Section~\ref{sec_challenge_desc}), provide an overview of submitted systems (Section~\ref{sec_systems}), dive deeper into multi-channel (MC) and single-channel (SC) tracks (Sections~\ref{sec_mc_systems} and \ref{sec_sc_systems}), discuss the advantages of MC over SC for DASR (Section~\ref{sec_mc_vs_sc}), and discuss approaches to ASR adopted by the submitted works. We conclude the paper by summarizing the key findings in Section~\ref{sec_conclusions}.

\section{NOTSOFAR-1 challenge description}
\label{sec_challenge_desc}

\subsection{Overview}
\label{sec_tracks}
The challenge featured two main tracks: the single-channel (SC) track and the known-geometry multi-channel (MC) track. Participants could submit entries to one or both tracks, allowing a comprehensive evaluation of their systems.

\textbf{Single-Channel:} This track focused on the use of a single recording device to capture audio in meeting scenarios. The goal was to evaluate DASR systems that can effectively handle the complexities of distant speech recognition using a single audio channel.

\textbf{Known-Geometry Multi-Channel:} This track utilized multiple recording devices with a fixed, known geometry (one central microphone and six surrounding microphones), aimed at evaluating the advantages of using spatial information from the MC setup to enhance DASR performance.

\subsection{Datasets}
\label{sec_datasets}
The challenge introduced two new datasets to the community--a recorded meeting dataset and a simulated training dataset. We describe the key attributes of each dataset below and refer the reader to \cite[Section 4]{vinnikov2024notsofar} for comprehensive details.\footnote{Further details of the datasets, including metadata analysis, are available on the challenge's website: \url{https://www.chimechallenge.org/challenges/chime8/task2/data}.}

\textbf{Recorded Meeting Dataset:} Comprises 315 unique English meetings with high-quality transcription, each of which averages six minutes in duration, recorded in 30 different rooms with both SC and MC devices. Each meeting features 4-8 adult participants, with 22 unique speakers participated in the development and training sets, and a separate set of 13 speakers in the evaluation set. 

The transcription process included a multi-judge annotation, and avoided machine assistance to minimize potential biases.
The recordings capture a broad spectrum of acoustic and conversational patterns, e.g., different topics, moving speakers, un/crowded meetings, overlap speech, etc., where part of this information is provided as metadata for error analysis. 

For completeness, Table~\ref{tab:meeting_ds} presents the summary table from \cite[Table 1]{vinnikov2024notsofar}, which outlines the technical details of this dataset.

\begin{table}[h]
    \centering
    \resizebox{0.9\linewidth}{!}{
        \begin{tabular}{|c|c|c|c|}
            \hline
            \textbf{Feature} & \textbf{Training Set} & \textbf{Development Set} & \textbf{Evaluation Set (blind)} \\
            \hline
            Number of Meetings & 110 meetings & 35 meetings & 170 meetings \\
            \hline
            Average Meeting Duration & 6 minutes & 6 minutes & 6 minutes \\
            \hline
            Number of SC streams & 5 streams & 5 streams & 6 streams \\
            \hline
            Number of MC devices & 4 devices & 4 devices & 4 devices \\
            \hline
            \makecell{Total Duration of SC \\ Audio Recordings} & \makecell{55 hours \\(110 meetings $\times$ \\ 6 minutes $\times$ \\ 5 SC streams)} & \makecell{17.5 hours \\ (35 meetings $\times$ \\ 6 minutes $\times$ \\ 5 SC streams)} & \makecell{102 hours \\ (170 meetings $\times$ \\ 6 minutes $\times$ \\ 6 SC streams)} \\
            \hline
            \makecell{Total Duration of MC \\ Audio Recordings} & \makecell{44 hours \\ (110 meetings $\times$ \\ 6 minutes $\times$ \\ 4 MC devices)} & \makecell{14 hours \\ (35 meetings $\times$ \\ 6 minutes $\times$ \\ 4 MC devices)} & \makecell{68 hours \\ (170 meetings $\times$ \\ 6 minutes $\times$ \\ 4 MC devices)} \\
            \hline
            Total Number of Rooms & 20 rooms & 5 rooms & 13 rooms \\
            \hline
            Number of Speakers & 22 speakers & 11 speakers & 13 speakers \\
            \hline
            Language & \multicolumn{3}{c|}{English} \\
            \hline
        \end{tabular}
    }
    \caption{A summarization of NOTSOFAR recorded meeting dataset as presented in \cite[Table 1]{vinnikov2024notsofar}. SC and MC stand for Single-Channel and Multi-Channel, respectively.}
    \label{tab:meeting_ds}
\end{table}

\textbf{Simulated Training Dataset:} Consists of approximately 1000 hours of simulated data, carefully crafted to effectively represent real-world speech conditions. It incorporates 15,000 \emph{real} ATFs, and features high-quality clean speech. This simulated dataset is designed to support the training of speech separation and enhancement methods, addressing the limitations of previous simulated datasets to close existing train-test gaps.

\subsection{Metrics}
\label{sec_metrics}
The evaluation of the submitted systems was based on two key metrics, described hereafter. These metrics were computed for each session and then averaged to provide a final score for each system. The results were used to rank the participating teams and identify the best performing system.

\textbf{Time-constrained
minimum-permutation word-error rate (tcpWER) \cite{neumann2023meeteval}:} A time-constrained version of the Concatenated minimum-Permutation WER (cpWER) \cite{watanabe2020chime6}. It measures word error rate while attributing errors to specific speakers within a time-constrained window, providing an assessment of both speech recognition and speaker diarization performance.

\textbf{Time-constrained optimal reference combination word-error rate
(tcorcWER) \cite{neumann2023meeteval}:} A time-constrained version of the Optimal Reference Combination WER (orcWER) \cite{sklyar2022multi}. It optimizes word error rate calculation by dynamically selecting the most accurate reference transcriptions within specific time windows, focusing solely on the accuracy of the transcribed text without considering speaker identity.

\subsection{CHiME-8 DASR and NOTSOFAR tasks}
\label{sec_dasr}
In addition to NOTSOFAR-1 task being summarized by this paper, CHiME-8 challenge also offered DASR task focusing on microphone-array geometry agnostic systems. Both tasks focus on distant automatic speech recognition and speaker diarization, offering a fundamental comparison among different system designs when evaluated on NOTSOFAR-1: geometry-agnostic (DASR) versus known geometry (NOTSOFAR-1). Since none of the NOTSOFAR-1 teams participating in the multi-channel track used the core training datasets for the DASR task (CHiME-6, DiPCo, Mixer6), we have included only NOTSOFAR-1 entries in the official ranking for a fair comparison. 
However, since we never intended to introduce any limitations on the training data and taking into account the remarkable performance of DASR submissions on NOTSOFAR-1 data, we decided to include them in the analysis for the sake of more objective review and better scientific insights. We suggest to keep in mind DASR submissions, including those ranked as second and third in the analysis below (STCON \cite{mitrofanov24stcon} and NTT \cite{kamo24ntt}), are the only systems that used DASR training data in addition to NOTSOFAR-1 training data.

\subsection{Results}
We received 9 submissions in the single-channel track and 5 NOTSOFAR-1 submissions in the multi-channel track. We've also included 3 DASR track submissions to the multi-channel track (see \ref{sec_dasr} for more details). Most of the submissions improved the baselines, some very significantly. Tables~\ref{tab:sc_results} and \ref{tab:mc_results} show tcpWER and tcorcWER for all submissions to the SC and MC tracks respectively. 

\begin{table}[h]
    \centering
    \resizebox{0.9\linewidth}{!}{
        \begin{tabular}{|c|c|c|c|c|}
            \hline
            Rank & Team Name & System Tag & tcpWER (\%) & tcorcWER (\%) \\ \hline
            1 & USTC-NERCSLIP \cite{niu24ustc} & sys2 & 22.2 \% & 17.7 \% \\ \hline
            2 & NPU-TEA \cite{huang24npu} & sys4 & 30.0 \% & 25.7 \% \\ \hline
            3 & NJU-AALab \cite{hu24nju} & sys1 & 33.5 \% & 30.4 \% \\ \hline
            4 & NAIST \cite{hirano24naist} & sys2 & 36.6 \% & 33.3 \% \\ \hline
            5 & BUTJHU \cite{polok24but} & sys2 & 40.1 \% & 27.2 \% \\ \hline
            6 & ToTaTo \cite{kalda24totato} & sys4 & 41.2 \% & 29.2 \% \\ \hline
            7 & NOTSOFAR-1 baseline & sys1 & 41.4 \% & 35.5 \% \\ \hline
            8 & Fano Labs \cite{broughton24fano} & sys3 & 43.1 \% & 40.0 \% \\ \hline
            9 & UWB & sys1 & 45.8 \% & 39.1 \% \\ \hline
            10 & Blue Sky Wave Riders & sys2 & 74.1 \% & 37.0 \% \\ \hline
        \end{tabular}
    }
    \caption{Single-channel track results (NOTSOFAR-1 eval set)}
    \label{tab:sc_results}
\end{table}

\begin{table}[h]
    \centering
    \resizebox{0.9\linewidth}{!}{
        \begin{tabular}{|c|c|c|c|c|}
            \hline
            Rank & Team Name & System Tag & tcpWER (\%) & tcorcWER (\%) \\ \hline
            1 & USTC-NERCSLIP \cite{niu24ustc} & sys1 & 10.8 \% & 9.5 \% \\ \hline
            2 & STCON (DASR) \cite{mitrofanov24stcon} & sys1 & 13.8 \% & 11.6 \%  \\ \hline
            3 & NTT (DASR) \cite{kamo24ntt} & sys3 & 15.9 \% & 12.6 \%  \\ \hline
            4 & NPU-TEA \cite{huang24npu} & sys3 & 18.7 \% & 17.0 \%  \\ \hline
            5 & NAIST \cite{hirano24naist} & sys2 & 23.0 \% & 21.1 \%  \\ \hline
            6 & BUTJHU \cite{polok24but} & sys2 & 24.9 \% & 15.6 \%  \\ \hline
            7 & NOTSOFAR-1 baseline & sys1 & 28.3 \% & 24.6 \%  \\ \hline
            10 & Blue Sky Wave Riders & sys2 & 71.9 \% & 34.4 \%  \\ \hline
            11 & Anonymous Team (DASR) & sys1 & 85.3 \% & 59.2 \%  \\ \hline
        \end{tabular}
    }
    \caption{Multi-channel track results (NOTSOFAR-1 eval set)}
    \label{tab:mc_results}
\end{table}

\section{Overview of the submitted systems}
\label{sec_systems}

Transcription DASR pipelines typically start with a front-end module, followed by an automatic speech recognition (ASR) system. During the front-end stage, the input audio stream is processed to separate overlapping speech and reduce noise and reverberation. This step is essential as it prepares the audio for the ASR system, which then converts the enhanced audio into text.

In the NOTSOFAR-1 challenge, all submitted systems implemented one of two types of transcription pipelines, described below. In this paper we refer to these pipelines as CSS-ASR-Dia and Dia-Sep-ASR.

\begin{itemize}
    \item \textbf{CSS-ASR-Dia} pipelines start with continuous source separation (CSS) \cite{chen2020css} as a front-end, followed by automatic speech recognition (ASR) and speaker diarization (Dia). CSS module generates a set of non-overlapped speech signals from a continuous audio stream that contains multiple utterances that are partially overlapped by a varying degree. Speech signals produced by CSS are processed by ASR. Since the speech signals generated by CSS can combine multiple speakers -- CSS aims to ``unmix'' overlapping speech  rather than isolate individual speakers -- ASR is followed by speaker diarization. 
    \item \textbf{Dia-Sep-ASR} pipelines separate the tasks of audio segmentation and source separation. They start with a speaker diarization module that determines \emph{who spoke when} and returns a set of time intervals per detected speaker. The audio is then divided into intervals based on the diarization output, source separation being applied to the intervals containing overlapping speech. Lastly, an ASR module is applied to the separated sources for transcription. 
\end{itemize}

As challenge designers, we aimed to encourage participants to explore replacing traditional GSS-based approaches to source separation with data-driven neural-network alternatives. This led us to provide a CSS-ASR-Dia pipeline as the baseline system.
We also wanted to give participants the option to adapt the ASR to the distant speech as opposed to improving the acoustic front-end in order to bridge the gap between distant and close, so the ASR module was not fixed and participants were encouraged to come up with alternatives.

Some teams built upon the provided baseline system, focusing on enhancing its pipeline modules, while others opted for the more established Dia-Sep-ASR pipeline structure. Additionally, many participants invested in fine-tuning ASR systems to improve performance.

Tables \ref{tab:systems_and_technique_mc} and \ref{tab:systems_and_technique_sc} show the techniques used by MC and SC systems respectively. Participants could submit up to four systems, and the tables include all reported techniques, not just those from the best-performing solutions. Submissions that underperformed compared to the baseline on the evaluation set are excluded.

\begin{table}[h]
    \centering
    \resizebox{1.0\linewidth}{!}{
        \begin{tabular}{|p{2.5cm}|p{5cm}|p{3.5cm}|p{4cm}|p{3cm}|}
            \hline
            Submission & Diarization & Separation & 
            ASR and LM & 
            Post-ASR Dia 
            \\ \hline
            USTC- \newline NERCSLIP 
            & 
            Conformer CSS and overlap detection, \newline 
            Beamforming, \newline 
            ResNet-221, \newline 
            Spectral clustering, \newline
            NSD-MS2S, \newline 
            cACGMM Rectification, \newline 
            GSS 
            &
            JDS, \newline GSS, \newline MVDR 
            & 
            Enhanced Whisper 
            & 
            \\ \hline
            STCON 
            & 
            Amplitude normalization, \newline
            WPE, \newline
            MicRank channel selection, \newline  
            VAD, \newline
            Wav2Vec2.0 XLS-R 531, \newline
            ECAPA-TDNN, \newline
            UMAP dim. reduction, \newline
            DBSCAN, \newline
            GMM clustering, \newline
            NSD-MS2S, \newline
            DOVER-Lap 
            &
            GSS (WPE, \newline cACGMMs, \newline MVDR), \newline  
            G-TSE \newline
            &
            DNN-HMM, \newline
            Uconv-Conformer, \newline
            E-Branchformer, \newline
            WavLM, \newline
            Llama-2-7B \newline 
            & 
            \\ \hline
            NTT &
            Channel clustering, \newline
            EEND, \newline
            TS-VAD, \newline
            NSD-MS2S, \newline
            ECAPA-TDNN, \newline
            GSS, \newline
            NME-SC, \newline
            DOVER-Lap
            &
            Channel selection, \newline
            WPE, \newline
            GSS (cACGMM), \newline
            SP-MWF,
            &
            Whisper Large v3, \newline 
            Whisper Medium, \newline
            NeMo Transducer, \newline
            WavLM Transducer, \newline
            Transformer-LM, \newline
            &
            \\ \hline
            NPU-TEA & &
            Conformer CSS
            &
            Silero VAD, \newline
            Whisper Large v2, \newline
            3-gram LM
            &
            ResNet293 or \newline
            ECAPA-TDNN, \newline
            WavLM
            \\ \hline
            NAIST & &
            nara-WPE, \newline
            Conformer CSS
            &
            WavLM, \newline
            Zipformer-T, \newline
            RNN LM
            & 
            TitaNet, \newline
            NME-SC
            \\ \hline
            BUT/JHU 
            &
            Multi-channel WavLM, \newline EEND
            &
            GSS
            &
            Whisper-based \newline TS-ASR
            &
            \\ \hline

        \end{tabular}
    }
    \caption{Models and techniques used in the MC submissions}
    \label{tab:systems_and_technique_mc}
\end{table}

\begin{table}[h]
    \centering
    \resizebox{1.0\linewidth}{!}{
        \begin{tabular}{|p{2.5cm}|p{5cm}|p{3.5cm}|p{4cm}|p{3cm}|}
            \hline
            Submission & Diarization & Separation & 
            ASR and LM & 
            Post-ASR Dia 
            \\ \hline
            USTC- \newline NERCSLIP 
            & 
            Neural overlap detection, \newline 
            CSS, \newline 
            ResNet-221, \newline 
            Spectral clustering, \newline
            NSD-MS2S
            &
            JDS
            & 
            Enhanced Whisper 
            & 
            \\ \hline
            NPU-TEA & &
            Conformer CSS, \newline
            WavLM
            &
            Silero VAD, \newline
            Whisper Large v2, \newline
            3-gram LM
            &
            ResNet293, \newline
            ECAPA-TDNN, \newline
            WavLM
            \\ \hline
            NJU-AALab
            &
            &
            TF-GridNet CSS, \newline
            Energy-based SPP
            &
            Whisper Large v2, \newline
            unspecified LM 
            &
            TitaNet, \newline
            multi-scale NME-SC, \newline
            k-means
            \\ \hline
            NAIST & &
            nara-WPE, \newline
            Conformer CSS
            &
            WavLM, \newline
            Zipformer-T, \newline
            RNN LM
            & 
            TitaNet,
            NME-SC
            \\ \hline
            BUT/JHU 
            &
            Multi-channel WavLM, \newline EEND
            &
            &
            Whisper-based \newline TS-ASR
            &
            \\ \hline
            ToTaTo
            &
            &
            PixIT CSS
            &
            Whisper large-v3
            &
            w2v-BERT 2.0, \newline 
            ECAPA-TDNN, \newline
            PixIT, \newline
            TitaNet, \newline
            NME-SC
            \\ \hline
        \end{tabular}
    }
    \caption{Models and techniques used in the SC submissions}
    \label{tab:systems_and_technique_sc}
\end{table}

\section{Multi-channel systems}
\label{sec_mc_systems}

\subsection{Top multi-channel systems}
The multi-channel track was won by USTC\cite{niu24ustc} team that employed the Dia-Sep-ASR approach (see Table~\ref{tab:mc_results}). The second and third places were secured by  STCON\cite{mitrofanov24stcon} and NTT\cite{kamo24ntt}, both of which competed in the DASR track aimed at microphone array geometry-agnostic solutions. Since STCON and NTT also utilized Dia-Sep-ASR pipeline structure, it is evident that this paradigm dominates the multi-channel track. 

All three top systems employed GSS \cite{boeddeker2018gss} as the core component of their source separation subsystems, reinforcing GSS's position as the leading approach for source separation.
While USTC's winning system incorporated CSS in its diarization subsystem, we still classify it as Dia-Sep-ASR since GSS handled the source separation later in the pipeline.
Notably, USTC also proposed an alternative pipeline that replaced GSS with a neural-network based joint diarization and separation (JDS) component, but their best-performing pipeline still employed GSS. We discuss this in greater detail in Section~\ref{beyond_gss}.

It is important to highlight that CSS-based pipelines are relatively new. Judging their currently lower performance compared to the well-established GSS-based systems as a lack of potential would be premature. In particular, we hope that CSS fine-tuning with real meeting data (a method unexplored by participants) holds promise for significantly improving the accuracy of CSS-based systems.

\subsection{Diarization and ASR}
All three top systems demonstrated significant investments in enhancing all stages of the transcription pipeline, with diarization and ASR receiving the most attention due to their apparently critical roles. 

GSS depends on accurate speaker diarization, so all three teams developed  sophisticated diarization subsystems combining clustering-based speaker diarization (CSD), neural speaker diarization (NSD), and speech separation components to refine speaker embedding in regions with overlapping speech. All leading diarization systems incorporated TS-VAD \cite{medennikov2020tsvad} techniques, specifically leveraging NSD-MS2S \cite{yang2024nsd_ms2s}. 
Given the importance of precise speaker counting for TS-VAD initialization, teams implemented advanced speaker-counting modules. STCON used a combination of an in-house version of Wav2vec 2.0 \cite{mitrofanov24stcon,baevski2020wav2vec} trained to extract several embeddings for segments with mixed speech and SpeechBrain ECAPA-TDNN model \cite{dawalatabad2021ecapa}. NTT speaker counting module employed ECAPA-TDNN speaker embeddings derived from on the audio processed by GSS initialized with EEND-based segmentation \cite{kinoshita2021eend,kamo24ntt}. USTC system ran CSS on the overlapping speech segments before calculating speaker embeddings \cite{niu24ustc}.

ASR modules of the top system also received substantial upgrades. USTC introduced ``Enhanced Whisper'' leveraging features extracted from self-supervised pre-trained models like WavLM \cite{chen2022wavlm}. They also introduced additional improvements like augmentation of the original Whisper model \cite{radford2023whisper} with a ConvNeXt structure \cite{jiang2022nextformer_convnext}, running in parallel to the standard 1D convolutions, adopted bias relative positional encoding, integrated sigmoid gating mechanism augmented the final layer of the encoder with a Mixture of Experts (MoE) component.
STCON used an ensemble of DNN-HMM hybrid models and end-to-end (E2E) models trained on GSS-processed CHiME-8 train datasets and the clean part of LibriSpeech \cite{panayotov2015librispeech} distorted with the simulated room impluse responses (RIR) selected to match those detected in the CHiME-8 training data.
NTT used an ensemble of fine-tuned Whisper Large v3, Whisper Medium, NeMo Transducer \cite{park23nemo} and fine tuned WavLM Transducer \cite{chen2022wavlm, kim2023branchformer}. The team also built a Transformer-LM used to rescore the n-best hypotheses of each model and combined the results using ROVER \cite{fiscus1997rover}. 

With all top teams investing heavily into multiple components, it is interesting to explore how the accuracy of each component affects the final speech recognition accuracy. While drawing decisive conclusions is impossible without thorough hands-on experimentation with the systems, the similar structure of the top systems and the ablation studies provided by the teams offer valuable insights into contribution of specific components into USTC's system success. 

Since all top systems relied on GSS for source separation and the improvements to the separation component proposed by the winning team (GSS initialization in T-F domain based on mask prediction by JDS \cite{niu24ustc}) didn't result in a major improvement based on the numbers reported by the team (Table 2 in \cite{niu24ustc}), we take the risk of assuming no significant difference in the systems source separation components and exclude the source separation from the discussion to focus on the accuracy of diarization subsystems. 

The ``Diarization'' column of Table~\ref{tab:systems_and_technique_mc} illustrates the complexity of diarization subsystems of top systems. USTC diarization subsystem \cite{niu24ustc} uses Conformer-based \cite{gulati2020conformer} overlap detector, separates overlapping speech with a Conformer-based CSS module to get cleaner speaker embeddings processed by the first round of clustering-based speaker diarization (CSD). Optionally, the first round of CSD is followed by multiple rounds of neural speaker diarization (NSD) and CSD, interspersed with speaker separation and speech enhancement modules like cACGMM rectification and GSS aiming at producing more refined speaker embeddings. 

However, despite the fact that every stage of their complicated diarization subsystem shows improvement in DER on dev-set-2 (Table 4 in \cite{niu24ustc}), the final tcpWER of the simplest system of all (system 2) is comparable with that of the rest of the systems (See Table 2 in \cite{niu24ustc}). Moreover, the best DER reported by the USTC team is significantly higher than that of other systems - 14.19\% on dev-set-2 (Table 4 in \cite{niu24ustc}) compared to 7.9\% reported by STCON (Table 2 in \cite{mitrofanov24stcon}, 9.72\% reported by NTT (Table 1 in \cite{kamo24ntt}) and 10.4\% reported by BUT/JHU (Table 1 in \cite{polok24but}). 

Given the observation that lower DER does not necessarily translate to lower tcpWER, and the inferior diarization accuracy of the winning system, we hypothesize ASR improvements offered by USTC played a key role in their system achieving the best tcpWER in both single-channel and multi-channel tracks. 

A possible explanation of this phenomenon lies in the goal of diarization: to segment the audio for GSS processing, determine the number of speakers in each segment and provide speech activity intervals per speaker for GSS initialization. While accurate speaker counting is crucial, some inaccuracies in speaker activity intervals used for GSS initialization may not significantly affect the GSS output. This makes certain aspects of diarization quality less critical and renders DER a less reliable predictor of the final ASR accuracy.   

Disclaimer: DER numbers mentioned in this section are self-reported by participants and could be calculated using different tools with different parameters (the ``collar'' parameter of DER calculation being probably the most important one). Although, strictly speaking, this makes the reported DER numbers incomparable and thus renders the conclusions above unreliable, we chose to still include them hoping they might serve as a potential guidance to researchers developing the next generation of transcription systems. We leave it to researchers to confirm or reject these insights by their own experimentation.

\subsection{Spatial information for diarization}
In addition to improving the SNR in the captured audio, microphone arrays make it possible to utilize the spatial signal based on differences in the sound wave propagation paths to different microphones of the array resulting in different time of arrival of the sound. Well known signal processing methods were developed to either detect the direction of sound arrival or amplify the sound coming from a direction of choice while attenuating the sound coming from other directions.

Although every submitted multi-channel system used that spatial info for speech separation through GSS \cite{boeddeker2018gss} or Minimum Variance Distortionless Response (MVDR) beamformer \cite{higuchi2017mvdr}, most of the systems ignored it for the diarization task and processed the audio channels independently benefiting only from aggregating multiple versions of diarization.  

BUT/JHU modified the WavLM model they used for diarization to enable comparison and exchange of information across channels with the goal to help model with extraction of important clues for diarziation, such as direction of arrival, which resulted in DER decrease from 10.9\% (single-channel) to 10.4\% (multi-channel) \cite{polok24but}. 

We believe the existence of well-established neural net based approaches to speaker diarization based on processing of a single-channel audio (EEND-VC \cite{kinoshita2021eend}, TSVAD \cite{medennikov2020tsvad}, NSD-MS2S \cite{yang2024nsd_ms2s}) together with the lack of such well established approaches for multi-channel audio is the reason for the dominance of single-channel oriented diarization subsystems in the challenge submissions. 
We hope that better use of the spatial signal offered by microphone arrays still presents an opportunity for a further significant improvement in multi-channel speaker diarization.

\subsection{Beyond GSS}
\label{beyond_gss}
Although GSS \cite{boeddeker2018gss} was a central source separation component of most of multi-channel systems, some teams tried to enhance the separation quality provided by GSS. USTC tried initialization of GSS in time-frequency domain, TF-masks for initialization being predicted by their neural joint diarization and separation module (JDS) \cite{niu24ustc}. STCON's systems include a neural target speaker enhancement (TSE) component serving as a ``GSS enhancer'' \cite{mitrofanov24stcon}. Also, one of the USTC systems separates the sources by MVDR beamformer \cite{higuchi2017mvdr} calculated based on T-F maps predicted by JDS, giving a hint of what is achievable without GSS \cite{niu24ustc}. 

\paragraph{Time-frequency domain initialization for GSS}
Table 2 of USTC paper \cite{niu24ustc} presents tcpWER for different combinations of diarization and separation subsystems.
The results show that the separation subsystem with T-F domain initialization of GSS (V2) performed only marginally better than the same subsystem without T-F domain initialization (V1) across all diarization front-ends. This suggests that T-F domain initialization of GSS doesn't significantly improve accuracy.
However, beyond accuracy, T-F domain initialization may reduce the number of iterations required for GSS optimization to achieve the desired separation quality. Unfortunately, the USTC team didn't report the number of iterations used ini their systems, making such comparison impossible.  

\paragraph{Multi-channel source separation without GSS}
V3 separation system by USTC replaces GSS-based source separation with MVDR beamformer \cite{higuchi2017mvdr} calculated based on by per speaker T-F masks predicted by a neural JDS module \cite{niu24ustc}.  
Comparing the performance of V1, V2 and V3 systems in Table 2 of the USTC report \cite{niu24ustc} provides insight into the ability of neural T-F mask prediction to replace GSS. V3 is consistently worse than both V2 (including both JDS and GSS) and V1 (GSS only), with the relative degradation of 3 to 4\% depending on the diarization.   
USTC team's decision to use the predicted T-F masks to guide a beamformer rather than use them directly to reconstruct the source audio suggests the ASR accuracy on directly reconstructed masks was inferior. If this assumption is true, STCON might observe a similar phenomenon of higher recognition error rate on neurally separated speech in spite of high quality of the restored audio \cite{mitrofanov24stcon}. STCON's discussion on Target Speaker Extractor (TSE) can provide some insight into the phenomenon. We provide the summary of that discussion in the following paragraph.

\paragraph{Target speaker extraction to replace linear beamforming}
\label{g-tse}
Another potential approach to improvement of the classical GSS+beamforming source separation is replacing a linear beamformer with a neural Target Speaker Extraction (TSE) model. This approach was explored by STCON team \cite{mitrofanov24stcon}. They used the conformer architecture offered by the baseline CSS model, but extended it to process target speaker info together with mixture. According to the team's report, standalone training of the TSE model didn't yield better ASR accuracy than GSS+beamforming pipeline despite the good quality of target speaker’s speech and high SNR in most of the TSE outputs.
The team attributed the gap in ASR accuracy to two factors: a mismatch between loss functions in TSE (si-sdr) and ASR (CTC+attention) and the longer context used by GSS compared to TSE. 
The first factor was addressed by concatenating TSE and ASR models and fine-tuning the pre-trained TSE model using the ASR loss. The second factor was tackled by utilizing the GSS output as a reference channel in the TSE output, effectively turning the TSE into a ``GSS signal enhancer'' which the team referred to as the Guided Target Speaker Extractor (G-TSE).
These two changes helped bridge the gap and achieve improvements over the GSS+beamforming pipeline, supporting the team's hypothesis.      

\paragraph{Target speaker ASR}
A notably different system was suggested by BUT/JSU \cite{polok24but}. The team explored the option of simplifying the pipeline and skipping the source separation altogether by conditioning their ASR model on frame-level diarization output and training it to recognize only target speech ignoring possible interference. The ASR module is called multiple times on the time intervals featuring multiple speakers to transcribe speech of each of them. The pipeline can be viewed as a variation of Dia-Sep-ASR pipeline where source separation and ASR are combined into a single module. Although the speaker-aware WER (tcpWER) of BUT/JHU system turned out to be inferior to other systems, its speaker-agnostic WER (tcorcWER) was better than that of any CSS-based system on the multi-channel track and worse than that of only one CSS-based system on the single-channel track (NPU-TEA \cite{huang24npu}).

\subsection{CSS-based systems}
\label{css_based_systems}
CSS-based systems like those submitted by NPU-TEA \cite{huang24npu} and NAIST \cite{hirano24naist} teams improved the baseline system by a lot, but failed to achieve the same transcription accuracy as Dia-Sep-ASR systems. The best tcpWER on the evaluation dataset achieved by a Dia-Sep-ASR system was 18.7\% (NPU-TEA) while the winning system achieved tcpWER=10.8\% (USTC). The baseline system tcpWER is 28.3\%.

Both NPU-TEA and NAIST used conformer architecture proposed by the baseline.   
The NAIST team reused the baseline CSS model, but added a de-reverberation module. Despite the CSS model being trained to predict reverberation-free signals, the NAIST team discovered reverberation in the model output. Their experiments shown that the added Weighted Prediction Error (WPE) module was responsible for reducing tcpWER from 30.3\% to 28.9\% on dev-set-2 (Figure 4 in \cite{hirano24naist}.)         

\paragraph{Fine-tuning CSS with real data} Although teams extensively used the provided real training data for fine-tuning their diarization and ASR model, the creators of the CSS-based systems relied solely on training with simulated data, leaving fine-tuning of CSS with real meetings data unexplored. Although, as the challenge designers, we invested significant efforts into realistic simulation of far-field audio and natural meeting dynamic, we believe the remaining gap between synthetic and real data may have prevented CSS from reaching its full potential. We encourage further exploration of CSS fine-tuning with real data and hope that bridging the data gap will unlock to the full capabilities of neural network-based CSS. We discuss this and other factors that could affect the performance of CSS-based system in the challenge in \ref{dia_sep_asr_vs_css_asr_dia}.

\section{Single-channel systems}
\label{sec_sc_systems}

Single-channel speaker diarization and source separation is more challenging due to the absence of spatial information, leaving only voice characteristics as the signal for the models to utilize.  

GSS \cite{boeddeker2018gss} used by most multi-channel systems for speech separation is based on spatial information and thus unavailable in the single-channel setting. Without GSS, most single-channel submitted systems adopted CSS-ASR-Dia pipeline letting the CSS module handle the source separation task. Participants explored various neural net architectures for CSS: conformer \cite{gulati2020conformer} (suggested by the baseline system), TF-GridNet \cite{wang2023tf_gridnet} and PixIT \cite{kalda2024pixit}. 

Two systems explored different approaches. 

\paragraph{Neural speech separation}
The winning system by USTC \cite{niu24ustc} is the only single-channel Dia-Sep-ASR system employing a neural joint diarization and separation model (JDS) predicting TF-mask for each speaker identified by the preceding diarization subsystem.
To get the separated audios for each speaker, the amplitude spectral features of the original mixed audio are multiplied by the T-F masks, and an inverse STFT transformation is performed.  

\paragraph{Target speaker ASR}
BUT/JHU \cite{polok24but} explored the alternative ``Target Speaker ASR'' approach where an ASR model is conditioned on frame-level diarization output, resulting in a simpler pipeline bypassing the speech separation stage. While the speaker-aware WER (tcpWER) of the BUT/JHU system was inferior to other systems, its speaker-agnostic WER (tcorcWER) outperformed all CSS-based system in the multi-channel track and was surpassed by only one CSS-based system in the single-channel track (NPU-TEA \cite{huang24npu}).

\section{Dia-Sep-ASR vs CSS-ASR-Dia}
\label{dia_sep_asr_vs_css_asr_dia}
In the previous sections we've concluded that Dia-Sep-ASR systems outperformed CSS-ASR-Dia systems in the challenge. In this section, we suggest several possible explanations to these observations. Without thorough ablation study of the systems, it is impossible to validate each of the offered explanations or estimate their relative impact, so we offer these potential explanations as hypotheses and encourage researchers to explore them in their future work.

While some observations apply to both SC and MC systems, the advantage of Dia-Sep-ASR systems was more evident in the MC track. The top three MC systems -- USTC \cite{niu24ustc}, STCON \cite{mitrofanov24stcon} and NTT \cite{kamo24ntt} -- are all Dia-Sep-ASR systems, demonstrating the effectiveness of this approach. The SC track was also won by a Dia-Sep-ASR system (USTC \cite{niu24ustc}); however, the absence of other high-performing Dia-Sep-ASR systems in this track (as STCON and NTT participated only in the DASR track and did not submit SC versions) makes the conclusion about Dia-Sep-ASR superiority less certain. Furthermore, the presence of USTC Enhanced Whisper \cite{niu24ustc} which likely contributed to USTC's performance, further complicates direct comparisons.

\subsection{Hypothesis 1: CSS is a more difficult task}
Defined as a process of generating a set of overlap-free speech signals from a continuous audio stream consisting of multiple utterances spoken by different people \cite{chen2020css}, the CSS task inherently combines speaker diarization and speech separation. The baseline and submitted CSS-ASR-Dia systems tried to tackle both tasks using a single CSS model. In contrast, high-performing Dia-Sep-ASR systems employed complex multi-model diarization and separation subsystems. For example, USTC diarization subsystem alone incorporated two modified baseline CSS models -- one repurposed for overlap detection and the other dedicated to speech separation for the sake of better speaker embeddings \cite{niu24ustc}. 

\subsection{Hypothesis 2: longer context windows may have contributed to the superior performance of Dia-Sep-ASR systems}
The baseline CSS model processed three-second speech segments, and submitted CSS-ASR-Dia systems maintained a similar window length. In contrast, top Dia-Sep-ASR systems incorporated components that  processed much longer windows, potentially leveraging context more effectively. GSS, for instance, commonly operates on windows significantly longer than three seconds. Even STCON's specialized ``chunked'' version of GSS, designed to accommodate frequent speaker movement, processed windows of about five seconds \cite{mitrofanov24stcon}. USTC's  T-F mask prediction network (JDS) and  overlap detection component of their diarization subsystem operated on 12.8-second windows, while their cACGMM rectification component used a 120-second window. These differences suggest that CSS performance could improve with access to longer context windows.

\subsection{Hypothesis 3: Dia-Sep-ASR systems may have been more effective in bridging gaps between simulated and real-world data} 
Despite significant efforts to make simulation data for model training as realistic as possible, some gaps between simulated and real data likely remain. We hypothesize that the submitted Dia-Sep-ASR systems may have been more effective in bridging these gaps by relying on GSS for speech separation, as it is less data-driven than neural network models. While neural networks generally benefit from better data utilization, simpler models like GSS may offer greater robustness when training and test data do not fully align.
As discussed in Section \ref{css_based_systems}, the submitted CSS models were trained solely on synthetic data. We believe that incorporating real-world data into CSS training could enhance its effectiveness and unlock the full potential of neural networks for speech separation.

\section{Single-channel vs multi-channel accuracy}
\label{sec_mc_vs_sc}
NOTSOFAR-1 challenge has confirmed the multi-channel systems' ability to provide significantly more accurate transcription in challenging settings compared to single-channel systems. The winning multi-channel system by USTC \cite{niu24ustc} achieved relative improvement of 51\% in terms of  speaker attributed word error rate (tcpWER) and 46\% in terms of speaker-agnostic word error rate (tcorcWER) over the winning single-channel system, also by USTC \cite{niu24ustc}. 
\begin{table}[h]
    \centering
    \resizebox{0.9\linewidth}{!}{
        \begin{tabular}{|c|c|c|c|}
            \hline
            Track & System & tcpWER, \% & tcorcWER, \% \\ \hline
            Single-channel & USTC-NERCSLIP, sys1  & 10.8 & 9.5 \\ \hline
            Multi-channel & USTC-NERCSLIP, sys2 & 22.2 & 17.7 \\ \hline
        \end{tabular}
    }
    \caption{The accuracy of the winning single-channel and multi-channel systems}
    \label{tab:single_vs_multi}
\end{table}
\subsection{The effect of adversarial factors}
\begin{figure}[h!]
    \centering
    \begin{minipage}{0.45\textwidth}
        \includegraphics[width=\textwidth]{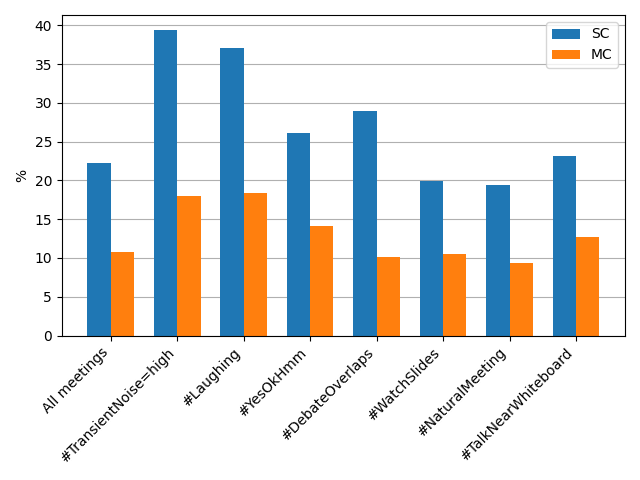}
        \caption{Best tcpWER per hashtag in SC and MC tracks}
        \label{fig:sc_mc_per_hashtag}
    \end{minipage}
    \hfill
    \begin{minipage}{0.45\textwidth}
        \includegraphics[width=\textwidth]{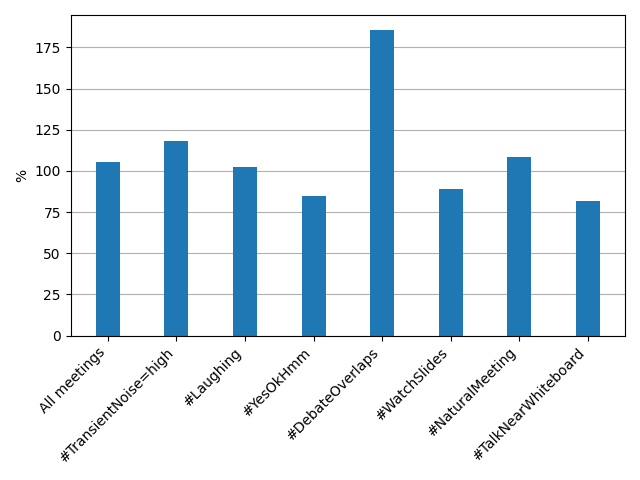}
        \caption{Relative increase of best tcpWER in SC track compared to MC per hashtag}
        \label{fig:sc_increase_per_hashtag}
    \end{minipage}
    
\end{figure}

Per-meeting hashtags in NOTSOFAR-1 dataset identify meetings with certain acoustic conditions or speakers behavioral patterns, enabling the analysis of how certain adversarial factors affect the performance of both SC and MC systems. Figure~\ref{fig:sc_mc_per_hashtag} presents the best tcpWER scores achieved by SC and MC systems across all evaluation set meetings, as well as for meetings tagged with various hashtags. Figure~\ref{fig:sc_increase_per_hashtag} illustrates the relative increase in tcpWER between the best SC and best MC systems.

Transient noise and laughter have the most significant impact on the performance of both SC and MC systems. 

Debate-style overlapping speech is particularly challenging for SC systems, but MC systems can handle it effectively, achieving accuracy comparable to their average performance on all meetings or on ``natural'' meetings, where participants were not given specific instructions to create additional challenges. 
The success of MC systems in dealing with debate-style overlapping speech can be attributed to the advantages of using spatial signal, which enables source separation even during rapid turns and complex conversation dynamics. In contrast, SC systems, relying solely on voice characteristics, struggle with these conditions.

\#TalkNearWhiteboard hashtag marks meetings where at least one participant was standing next to the whiteboard while speaking. Figure~\ref{fig:sc_mc_per_hashtag} highlights the tcpWER increase for both SC in MC systems in such meetings. While the impact is less severe compared to some other adversarial factors, the tcpWER increase for speech from speakers standing next to the whiteboard is likely more pronounced.
Among the factors that make next to the whiteboard position more challenging for ASR are potentially larger distance to the microphones, but also a potentially unusual ATF caused by the reflective surface of the whiteboard close to the speaker mouth. If speakers write on the whiteboard while speaking, their face is also likely to be turned towards the whiteboard and away from the microphones, eliminating the direct path of the sound wave from the speaker mouth to the microphones.

\section{ASR}
\label{sec_asr}

\subsection{OpenAI Whisper}
OpenAI Whisper \cite{radford2023whisper} was offered as a part of the baseline system and proved to be extremely popular among challenge participants, featuring in most submitted systems. 
Despite Whisper's widespread adoption, including serving as the base for the "Enhanced Whisper" used by the winning team \cite{niu24ustc}, the second and third places in the multi-channel track were claimed by the systems employing ensembles of multiple ASR models (STCON \cite{mitrofanov24stcon}, NTT \cite{kamo24ntt}). Notably, the ensemble being used by the second place system didn't include Whisper. In the single-channel track the top three systems all utilizing fine-tuned versions of Whisper (USTC \cite{niu24ustc}, STCON \cite{mitrofanov24stcon}, NTT \cite{kamo24ntt}). The winning system featured "Enhanced Whisper" \cite{niu24ustc}, which incorporated modifications to the model architecture and leveraged WavLM-based \cite{chen2022wavlm} features for further improvements.

\subsection{ASR model adaptation}
ASR model adaptation proved to be highly effective. All participating teams invested in adapting ASR models to far-field audio preprocessed with speaker separation (CSS or GSS) and reported significant improvements compared to ``vanilla Whisper'' baseline. 

For instance, Table 8 in the USTC paper \cite{niu24ustc} demonstrates a tcpWER reduction from 16.57\% to 9.87\% (a relative improvement of 40\%) on Dev-set-2 after incorporating multi-channel audio from Train-set-1/2, Dev-set-1 and LibriSpeech \cite{panayotov2015librispeech} simulated data processed with oracle GSS into the training set. 

NTT \cite{kamo24ntt} identified challenges in ASR adaptation due to some of CHiME-8 training data being very noisy and thus unreliable for fine-tuning. As a solution, they implemented curriculum learning where utterances with high character error rate (CER) were initially excluded from training by changing the target to the self-generated decoding and weighting the loss by 0.001. As the adaptation progressed and CER decreased, more difficult utterances were gradually reintroduced into the training process.

NPU-TEA \cite{huang24npu} adopted a similar iterative approach to fine-tune Whisper-large-v2. Similarly to NTT's method, utterances with too high edit distances to the ground truth (GT) were initially excluded, but could be included again in later iterations if their edit distance to GT improved as a result of the adaptation process.

\subsection{Inference techniques}
In addition to training time techniques such as model adaptation to source separation outputs and models architectural improvements, participants introduced inference time techniques that further improved Whisper's performance. The techniques included:
\begin{itemize}
    \item Filtering out non-speech segments to prevent hallucinations (NJU-AALab \cite{hu24nju}).
    \item Providing Whisper with longer context, achieved by merging short speech segments into longer ones (NJU-AALab \cite{hu24nju}) or using the previous decoding history as a decode prompt (USTC \cite{niu24ustc}).
    \item Prompting Whisper to return verbatim transcription (NJU-AALab \cite{hu24nju}). Without such prompting Whisper tends to correct speech disfluencies (e.g., repetitions, corrections, filler words), which lowers tcpWER since the ground truth retains all spoken words unchanged. 
\end{itemize}

Table 1 of the NJU-AALab paper \cite{hu24nju} shows that filtering out empty segments, merging short speech segments and prompting Whisper to return verbatim transcript together reduced tcorcWER on the development set from 37.6\% to 31.0\% (an impressive reduction of 17.5\%).   

\subsection{Re-scoring ASR output with language models (LM)}
Several teams employed rescoring n-best ASR results using fine-tuned LM models, which ranged from traditional 3-gram models to large language models (LLMs).

STCON utilized the Llama-2-7B LLM\footnote{https://huggingface.co/meta-llama/Llama-2-7b-hf}, fine-tuned on transcripts from CHiME-8 training data augmented with LibriSpeech texts. During inference, a context of 512 tokens, derived from 1-best results of previous utterances, was used. LLM rescoring achieved relative tcpWER reduction of 1.42\% to 2.94\% across different CHiME-8 datasets, with a reduction of 1.87\% (from 19.07\% to 18.72\%) on the NOTSOFAR-1 dev set.

NTT \cite{kamo24ntt} built a Transformer-LM with 35M parameters and a vocabulary of 1000 BPE tokens for LM rescoring. This model was pre-trained on 10\% of the LibriSpeech \cite{panayotov2015librispeech} text dataset and fine-tuned on the CHiME-8 training text dataset. At the inference, it used 256 previously rescored 1-best tokens as context.

NPU-TEA \cite{huang24npu} calculated the perplexity of the n-best results using a 3-gram LM, selecting the final recognition result by weighting and summing the perplexity and posterior probabilities. Their 3-gram LM rescoring was a part of their best multi-channel system. Table 2 of their paper showed their best system (A4) achieving 1.6\% improvement in tcpWER compared to A2, which didn't include rescoring. However, tcORC WER of A2 and A4 systems remained nearly identical (20.50\% vs 20.46\%). 

NAIST \cite{hirano24naist} observed that shallow fusion of ASR scores with those of RNN LM consisting of three layers of 2084-dimensional LSTMs \cite{hochreiter1997lstm} trained on LibriSpeech \cite{panayotov2015librispeech} dataset slightly reduced tcpWER achieved by their system. 

The winning system by USTC \cite{niu24ustc} didn't use LM-based rescoring in their pipeline.

\section{Conclusions}
\label{sec_conclusions}
This paper presented an overview of the NOTSOFAR-1 challenge, highlighting key results, insights and ideas for future improvement in DASR. The challenge captured the attention of the research community by introducing a dataset tailored not only to real-world scenarios but also to the pressing demands of business applications. With its combination of 315 recorded meetings across 30 distinct rooms and a simulated training dataset of 1,000 hours incorporating 15,000 real acoustic transfer functions (ATFs), the challenge set a new benchmark for realistic and diverse data in the field. This comprehensive dataset, paired with an open-source baseline system, attracted significant engagement from researchers, resulting in 11 submitted technical reports (9 from NOTSOFAR-1 and 2 from DASR tasks) that explored innovative solutions and demonstrated the potential for advancing DASR technologies. By bridging the gap between existing benchmarks and real-world applications, NOTSOFAR-1 established itself as a pivotal initiative, fostering collaboration and pushing the boundaries of speech processing research. 

The challenge confirmed the superiority of Dia-Sep-ASR systems over CSS-ASR-Dia systems in both single-channel (SC) and multi-channel (MC) tracks. The geometry-agnostic systems demonstrated competitive performance, achieving accuracy surpassed only by the winning system.

In the multi-channel setting, GSS proved to still be the dominant technique for source separation, underscoring its continued importance despite attempts to enhance it with neural network components. Neural network based Joint Diarization and Separation (JDS) approach offered by USTC \cite{niu24ustc}, however, proved to be viable alternative, excelling in the single channel and challenging GSS-based systems in multi-channel track.

One of the clear takeaways was the critical role of adapting ASR models to the specific acoustic challenges presented by far-field audio. Real-world data played an important role in improving the accuracy of diarization and ASR systems, while the untapped potential of fine-tuning CSS with real data presents a valuable avenue for future exploration. Interestingly, most teams relied on established single-channel neural diarization methods, underutilizing the spatial information offered by multi-channel audio setups.

The results also highlighted that multi-channel systems significantly outperformed single-channel systems, especially in challenging acoustic conditions and complex conversational dynamics of realistic meetings. The significant relative improvement in transcription quality reinforces the importance of spacial information in improving DASR performance.

While the NOTSOFAR-1 challenge showcased substantial progress in DASR technologies, it also highlighted possible areas for future investigation, including joint training of frontend and ASR models, better utilization of spatial signals for diarization, and fine-tuning of the models based on real-world data. These insights not only provide a roadmap for improving future DASR challenges but also drive innovation in real-world applications of conversational speech recognition.

Ultimately, the NOTSOFAR-1 Challenge has served as a catalyst for advancing the state of the art in distant speech transcription, contributing valuable datasets and benchmarks to the research community. These efforts will continue to inspire new approaches and applications in the rapidly evolving field of speech processing.

\section{Acknowledgments}
We extend our heartfelt thanks to Dr. Samuele Cornell (Carnegie Mellon University) for his invaluable advice and guidance throughout this work, as well as his assistance in organizing the NOTSOFAR-1 Challenge. His support has been instrumental in advancing this initiative and fostering collaboration within the research community.

Additionally, we express our sincere gratitude to all participating teams for their dedication, innovative contributions, and for taking the time to answer our questions regarding their submissions. Your engagement and openness have greatly enriched the challenge and facilitated the comprehensive analysis presented in this work. Your efforts have been pivotal in advancing the field of distant automatic speech recognition.

\section{Declaration of generative AI and AI-assisted technologies in the writing process.}
During the preparation of this work the authors used OpenAI ChatGPT in order to improve the readability and language of the manuscript. After using this tool/service, the authors reviewed and edited the content as needed and take full responsibility for the content of the published article.

 \bibliographystyle{elsarticle-num} 
 \bibliography{bibliography}






\end{document}